\documentclass[reprint,aip,jcp]{revtex4-1}
\usepackage{amssymb}
\usepackage{graphicx}
\usepackage{array}  
\usepackage{hhline}
\usepackage{longtable}
\cleardoublepage

\renewcommand{\thetable}{\Roman{table}} \thetable

\begin{document}
\title{On the corrections to Strong-Stretching Theory for 
       end-confined, charged polymers in a uniform electric field}

\author{Gabriele Migliorini}

\affiliation{Department of Mathematics, University of Reading, \\
Whiteknights, Reading RG6 6AX, United Kingdom}

\begin{abstract}
We investigate the properties of a system of semi-diluted polymers in 
the presence of charged groups and counter-ions, by means of 
self-consistent field theory. We study a system of polyelectrolyte chains 
grafted to a similarly, as well as an oppositely charged surface, solving 
a set of saddle-point equations that couple the modified diffusion 
equation for the polymer partition function to the Poisson-Boltzmann equation 
describing the charge distribution in the system. A numerical study of this 
set of equations is presented and comparison is made with previous studies. 
We then consider the case of semi-diluted, grafted polymer chains in the 
presence of  charge-end-groups. We study the problem with self-consistent 
field as well as strong-stretching theory. We derive the corrections to 
the Milner-Witten-Cates (MWC) theory for weakly charged chains and show that the 
monomer-density deviates from the parabolic profile expected in the uncharged case. 
The corresponding corrections are shown to be dictated by an Abel-Volterra integral 
equation of the second kind. The validity of our theoretical findings is confirmed 
comparing the predictions with the results obtained within numerical self-consistent field theory.
\end{abstract}

\maketitle

\def\s{\rule{0in}{0.28in}}

\section{Introduction}
\label{SectionI}

The stability of a dispersion against coagulation has been one of the primary 
subjects of research in colloid science. Colloidal stability can be obtained 
in polar solvents by means of ionic molecules. In this case an electrical 
double layer forms at the interface between the colloidal particles and the 
solvent, and the dependence of its structure on the electrolyte properties 
has been the key ingredient to explain electrostatic-induced colloidal 
stability in the past \cite{Shaw}.
In practice however electrostatic stability is difficult to obtain, due to 
extreme sensitivity of the process to the electrolyte 
conditions. A viable alternative to electrostatic stabilization is to 
absorb or end-graft non-ionic polymers on the surface of colloidal 
particles  \cite{Vincent,Milner1}. 
The stabilization mechanism obtained in this case is steric in nature.
The theory of stabilization of colloids by adsorbed neutral polymers 
\cite{Alexander,deGennes} is one of the first problems where self-consistent 
field theory has been applied \cite{Dolan,Dolan2}.
Self-consistent field theory, introduced in the context of neutral 
polymer systems by Edwards \cite{Edwards} and Helfand \cite{Helfand}, 
together with its discrete, lattice formulation counterpart \cite{Scheutjens}, 
has been applied and extended to a variety of inhomogeneous systems, 
including block-copolymer systems \cite{Matsen1}, polymers at interfaces, 
as well as non-ionic polymer brushes \cite{Orland,Netz,Matsen2}.
The study of polymer chains terminally attached to an interface has been 
branching from the colloidal science area to many other domains of research. 
Polymer brushes are a topic of interest in surface science, 
in view of their ability to modify surface properties \cite{Taunton,Singh,Carignano}. 
In particular chain-end-functionalized polymer brushes, with 
uncharged \cite{Huang}, as well as charged \cite{Lahann,Gong,Carignano2} end-groups,  
have been used to design and control surface properties in a variety of 
different ways. Theoretical studies of charged-end-group brushes \cite{Zhulina3}, 
and their response to an electrical external field \cite{Andelman}, have been 
discussed recently.

Polyelectrolytes are macromolecules containing ionizable groups. 
Their tendency to dissociate in solution into charged groups and 
low-molecular-weight counter-ions makes polyelectrolytes an interesting 
problem, as both steric interactions of the high-molecular-weight structure 
and electrostatic interactions between counter-ions and charged groups, 
coexist at different length scales. 
Due to the intricate interplay of these different type of interactions, 
the properties of polyelectrolyte solutions, as well as the behavior of 
polyelectrolytes near an interface, have been considerably more difficult 
to explain than those of non-ionic polymers. 
Extensions of the lattice self-consistent field theory for neutral polymers 
\cite{Scheutjens} have been discussed \cite{Lyklema}. 
The self-consistent theory of inhomogeneous polyelectrolyte systems has 
been formulated \cite{Shi,Borukhov}, and recently reconsidered in the 
context of symmetric di-block polyelectrolytes and polyelectrolyte 
blends \cite{Wang,Wang2}.
As discussed in the past \cite{Pincus,Marcelja,Misra}, for a system 
of end-grafted polyelectrolytes, the charge distribution is localized at 
the interface and symmetry considerations, as well as considerations about 
the double layer that forms at the surface, simplify their study. 
We consider both the case of end-grafted polyelectrolyte chains and end-grafted polymer 
chains with charged-end-groups at a flat interface, by means of self-consistent field theory. 
Related studies have been presented in the past \cite{Muthukumar2,Orland2,Zhulina3,
Witte} and a detailed comparison of the results obtained within self-consistent 
field theory with the predictions of analytic strong-stretching theory \cite{Zhulina2} 
are discussed in this work. The conformational properties of the system in the presence of a 
uniform external field \cite{Andelman} are then addressed.
The response of a polyelectrolyte brush to an external electric field 
represents an important problem with several practical applications. 
Its role in the actuation of nano-size cantilevers, to cite one of 
the many possible applications, has been recently reported \cite{Huck}.  
We study a system of end-confined charged polymer chains and its 
response to a uniform external field. We discuss the conformational changes of 
the monomer-density and counter-ion distribution profiles, for both 
the case of a similarly and oppositely charged grafting surface. 
We propose a new mathematical framework, that generalizes 
strong-stretching theory for charged grafted polymers in the presence of 
a uniform field and we discuss the results of our findings. 

\section{Theory}
\label{SectionII}

We introduce in this section the theoretical methods to describe the 
properties of a polyelectrolyte system of chains attached to a flat interface. 
The system is composed of $n_p$ polymers chains of length $N_p$, grafted to a 
flat, charged substrate of area $ { \cal A}$ with grafting density 
$ \sigma= n_p / { \cal A}$, in the presence of a second surface, at a 
distance $d$ and in the presence of $n_c$ counter-ions of valence $z_c$ in solution.
We consider the case of strongly dissociating polyelectrolytes, so that the 
number and position of the charges along the polymer chains is fixed and 
determined by the ionization degree parameter $f$, the total charge per 
chain being $fN_p$. The charge distribution along the polymer will be 
referred as $z_p(s)$, $s$ being the parameter that measures distance along 
the polymer backbone. Both the uniformly charged as well as the neutral 
brush with charged-end-groups are described by the present formalism. 
The ionization degree, polymer length and grafting density determine the 
counter-ion concentration, according to the electro-neutrality condition, 
$n_cz_c+fN_p n_pz_p=0$. 
The presence of mono-valent salt may easily be included in the notation 
introduced below, as well as the presence of other 'small' and polymeric 
charged species, according to the general formulation \cite{Shi} discussed 
in Appendix \ref{APP:A}.
Each polymer occupies a volume $N_p/ \rho_0$ and has a natural end-to-end 
length of $aN_p^{1/2}$, $a$ being the Kuhn segment. We assume all charged 
species, namely counter- and co-ions, to be point-like particles, neglecting 
steric contributions due to non-polymeric species. 
A general formulation of the theory, where several species are present and 
the size of the counter-ions and co-ions is explicitly taken into account, 
is discussed in Appendix \ref{APP:A}.
The system is immersed in a solvent. As in the case of non-ionic systems,
the corresponding degrees of freedom can be integrated out, so that the 
effective interaction between the monomers is described by the effective, 
short-ranged excluded-volume interaction parameter $v$. 
The Bjerrum length $l_B=e^2/ \varepsilon ({ \bf r}) k_BT$, together with the 
reference density $ \rho_0$ and the excluded-volume parameter $v$,  
characterizes the properties of the system. 

We consider in this section a single polyelectrolyte species in the presence of counter-ions, so that $N_p=N$ 
and $n_p=n$.
The dimensionless counter-ion concentration is defined as
\begin{equation}  
\hat{ \phi}_c( { \bf r}) = \frac{1}{ \rho_0}  \sum_{\alpha=1}^{n_c} 
\delta \big ( { \bf r} - { \bf r}_c^{ \alpha} \big ),
\label{density1}
\end{equation}
and the monomer concentration is defined as
\begin{equation}  
\hat{ \phi}_p( { \bf r}) = \frac{N}{ \rho_0}  \sum_{\alpha=1}^{n} 
\int_0^1 \delta \big ( { \bf r} - { \bf r}_p^{ \alpha}(s) \big ) ds,
\label{density2}
\end{equation}
where the configuration of the $\alpha$-th chain, belonging to the polymer 
species $p$ is defined by the vector ${ \bf r}^{ \alpha}_p$ and where the 
position of the $ \alpha$-th counter-ion is given by ${ \bf r}_c^{ \alpha}$.

The partition function associated with a system of polyelectrolyte chains 
in the presence of counter-ions, tethered to a grafting surface with 
density $ \sigma$, can be converted in a statistical field theory 
\cite{Parisi}, according to the general formulation of 
model $F$ and model $K$, as defined in the classical inhomogeneous polymer 
system classification \cite{Fredrickson}.
After integrating out the solvent degrees of freedom, and considering the 
case of strong polyelectrolytes, so that the polymer charge distribution 
is independent on the location of each monomer and smeared along the chain 
with a uniform charge ratio $f$, the effective hamiltonian of the system reads

\begin{eqnarray}
&&~~~\beta { \cal H}[w_p,\psi] = \frac{1}{2} \int d { \bf r} \Big [ \frac{1}{v}  
w_p({ \bf r})^2+\frac{1}{4 \pi l_B }  | \nabla \psi( { \bf r})|^2 \Big ] \nonumber \\
&&- \sigma \int d { \bf r}_{ \perp} \ln { \cal Q}_p({ \bf r}_{\perp}, iw_p + iz_p 
\psi )- n_c \ln { \cal Q}_c ( i \psi), \\ \nonumber
\end{eqnarray}
where $w_p({ \bf r})$ and $ \psi({ \bf r})$ are the two fields related to monomer concentration 
and electrostatic potential, where $ { \cal Q}_p$ and $ { \cal Q}_c$ are the 
single-chain and counter-ion partition functions, defined and in Appendix ~\ref{APP:A} 
and where a uniform distribution for the grafting points is assumed. 
The excluded-volume parameter, also introduced in Appendix \ref{APP:A}, represents the 
effective interaction between monomers.
As both the electrostatic and monomer density fields are uniform in the $x$ 
and $y$ directions, parallel to the grafting surface \cite{Muller}, the 
single-chain polymer energy contribution, along the direction perpendicular 
to the grafting surface reads
\begin{equation}
\frac{ E [ z^{ \alpha};0,s]}{k_BT} = \int_0^s \Big [ \frac{3}{2 a^2N}| 
\dot{z}^{ \alpha} (t)|^2+w_p(z^{ \alpha}(t)) \Big ]dt.
\label{energy}
\end{equation}
The grafting density $ \sigma$ can be absorbed 
introducing rescaled polymer and counter-ion concentrations, 
\begin{equation}
\phi_p(z)= \frac{a \rho_0}{ \sigma N^{1/2}} \Big \langle \hat { \phi}_p(z) 
\Big \rangle ~~~
\end{equation}
and 
\begin{equation}
\phi_c(z) = \frac{a \rho_0}{ f \sigma N^{1/2}} \Big \langle \hat { \phi}_c(z) 
\Big \rangle, 
\end{equation}
that are now normalized according to 
\begin{equation}
\int_0^d \phi_p(z) dz = a N^{1/2}
\label{normalization1}
\end{equation}
\begin{equation}
\int_0^d \phi_c(z) dz= aN^{1/2}.
\label{normalization2}
\end{equation}
The corresponding field equations are 
\begin{eqnarray}
\label{field}
w_p(z) &=& \Lambda \phi_p(z)+ N_c \psi(z) \nonumber \\
w_c(z) &=& -N \psi(z), \\ \nonumber
\end{eqnarray}
where the reduced interaction parameter is defined as \cite{Matsen3},
\begin{equation}
\Lambda \equiv \frac{v \sigma N^{3/2}}{a \rho_0},
\end{equation}
and is related to the brush height $L$ according \cite{Kim1} to 
$L/aN^{1/2}=(4 \Lambda/ \pi^2)^{1/3}$, where we introduce the number of 
ions per polymer parameter $N_c=fN$, and where all length scales are 
expressed in units of $a N^{1/2}$.

\section{ Self-Consistent Field Theory}
\label{SectionIII}
The crux of self-consistent field theory is the one-dimensional polymer 
partition function,
\begin{eqnarray}
\label{pf}
&& q(z,z_0,s) = \int { \cal D} z_{ \alpha} 
\exp \Big ( - \frac{E [ z^{ \alpha};0,s]}{k_BT} \Big ) \nonumber \\
&&~~~ \times~~ \delta(z_{ \alpha}(s)-z) \delta(z^{ \alpha}(0)-z_0), \\ \nonumber 
\end{eqnarray}
for a section of $sN$ segments with its ends at $z^{ \alpha}(0)=z_0$ 
and $z^{ \alpha}(s)=z$.
The polymer partition function satisfies the modified diffusion 
equation \cite{Matsen1}
\begin{equation}
\label{spe1}
 \frac{ \partial}{ \partial s} q(z,z_0,s) = 
\Big [ \frac{a^2N}{6} \frac{ \partial^2}{ \partial z^2} 
- w_p(z) \Big ]q(z,z_0,s)
\end{equation}
where $q(z,z_0,0)=aN^{1/2}\delta(z-z_0)$ and $q(0,z_0,s)=0$, 
and where the field equations (\ref{field}) include a contribution from the 
dimensionless electrostatic potential $\psi(z)$, that satisfies 
a non-linear Poisson-Boltzmann equation
\begin{equation}
\frac{1}{L_B} \frac{ \partial^2}{ \partial z^2} \psi(z)= \phi_c(z)-\phi_p(z),
\label{spe2}
\end{equation}
with boundary conditions 
\begin{eqnarray}
\label{bc2}
\Lambda_{GC} \frac{ \partial}{ \partial z} \psi(z) \Big |_{z=0}&=&-1\nonumber \\
\Lambda_{GC} \frac{ \partial}{ \partial z} \psi(z) \Big |_{z=d}&=&+1,\\\nonumber
\end{eqnarray}
that describe an external surface charge, where
\begin{equation}
L_B \equiv [(ze)^2/\varepsilon a N^{1/2}k_bT]/ \sigma N_c
\end{equation}
is the rescaled Bjerrum length parameter and where the 
Gouy-Chapman length is defined as 
\begin{equation}
\Lambda_{GC} \equiv zf \sigma N / \Sigma L_B.
\end{equation}
The counter-ion distribution is given by
\begin{equation}
\phi_c(z)=\frac{a N^{1/2}}{{ \cal Q}_c} \exp ( - w_c(z)/N),
\end{equation}
where 
\begin{equation}
{ \cal Q}_c= \int_0^d \exp( - w_c(z)/N) dz.
\end{equation}
Once the solution $q(z,z_0,s)$ to the modified diffusion equation (\ref{spe1}) 
has been found, the concentration profile for a chain with its free-end 
at $z=z_0$ can be computed according to 
\begin{equation}
\phi_p(z;z_0)= \int_0^1 \frac{ q( \epsilon,z,1-s)q(z,z_0,s)}
{q( \epsilon,z_0,1)}ds,
\label{spe3}
\end{equation}
where $\epsilon$ is a small finite distance from the grafting surface 
introduced according \cite{Muller} to the standard formulation of model $K$.
The free-end distribution is defined as 
\begin{equation}
g(z_0)= \frac { aN^{1/2}}{ { \cal Q}_P} 
\exp \big ( - \frac{f_0(z_0)}{k_BT} \big ),
\end{equation}
where
\begin{equation}
f_0(z_0)=- \ln q(0,z_0,1)
\end{equation}
and where finally
\begin{equation}
{ \cal Q}_p= \int_0^d \exp \big ( - \frac{f_0(z_0)}{k_BT} \big ) dz_0,
\end{equation}
and the monomer density distribution satisfies 
\begin{equation}
\phi_p(z)= \frac{1}{ a N^{1/2}} \int_0^d g(z_0) \phi_p( z;z_0) dz_0.
\label{concentration}
\end{equation}
The set of self-consistent field equations (\ref{spe1})-(\ref{spe2}) 
have been solved numerically, 
for different values of the four fundamental parameters, namely the 
natural brush height $L/aN^{1/2}$, the number of ions per polymer 
$N_c\equiv fN$, the rescaled Bjerrum Length $L_B$ and the relative surface 
charge parameter 
\begin{equation}
f_e=\Sigma/ zf \sigma N,
\end{equation}
where $\Sigma$ is the total charge on the grafting surface in elementary 
charge units. 
The free-energy of the system can be computed as 
\begin{eqnarray}
\label{fe}
&&~~~\frac{F}{k_BTn_c}= \frac{1}{ aN^{1/2} N_c} \int_0^{d} g(z_0) 
\big [ \frac{f_e(z_0)}{k_BT}+\ln g(z_0) \big ] dz_0 \nonumber \\
&& + \frac{1}{aN^{1/2}} \int_0^{ d} dz \phi_c(z) \ln \phi_c(z) dz + \frac{ \Lambda}{ 2 aN^{1/2}N_c} \int \phi_p^2(z)dz \nonumber \\
&& ~~~~~~~~~~+ \frac{1}{2 aN^{1/2}} 
\int_0^{d} \psi(z) \big [ \phi_p(z)-\phi_c(z) \big ]dz, \\ \nonumber 
\end{eqnarray}
where the first term is the polymer free-energy and is obtained as the sum 
of the translational entropy of the chain ends and the average of the 
stretching energy
\begin{equation}
\frac{f_e(z_0)}{k_BT}=- \ln q( \epsilon,z_0,1)- \frac{1}{ aN^{1/2}} 
\int w(z) \phi(z,z_0) dz, 
\label{feb}
\end{equation}
the second term represents the translational entropy of the mobile 
counter-ions and the third and fourth terms the interaction energy, due to 
excluded-volume interactions between polymer segments and the electrostatic 
energy related to the charge distribution in the system respectively. 
Details on the derivation of equation (\ref{fe}) can be found in 
Appendix \ref{APP:B}.

\section{Strong-stretching theory}
\label{SectionIV}
When the polymer chains are stretched it is possible to exploit the 
analogy between polymer field theory and quantum mechanics and approximate 
the single-chain partition function in equation (\ref{pf}), defined above within the 
path-integral formalism, by considering the dominant contribution of the 
classical path and the fluctuations around it \cite{Milner2,Johner}. 
The problem in this limit relates to the classical mechanics problem of a 
particle moving down an incline \cite{Matsen2}.
In the classical limit, the partition function $q(0,z_0,1)$ is dominated by 
the path $z_{\alpha}(s)$ that minimizes the energy in equation (\ref{energy}) and an 
estimate for the free-energy $f_0(z_0)$ of a chain extending to $z=z_0$, can 
be obtained accordingly:
\begin{equation}
q(0,Z_0,1) \propto \exp \Big ( - E[ Z^{ \alpha};1]/k_BT] \Big ), 
\end{equation}
where, from here on, we will use dimensionless units $Z=z/aN^{1/2}$.
Following the general principle of strong-stretching theory \cite{Milner2,Zhulina} 
we write, 
\begin{equation}
\frac{3}{2}S^2(Z,Z_0) \equiv N_c \big [ \frac{3}{2} v^2(Z_0)+U(Z)-U(Z_0) \big ],
\end{equation}
where the potential $U(Z)$ is to be identified with the polymer field $w_p(z)$ introduced in 
equation (\ref{field}). The chain-end tension can be written \cite{Johner} as
\begin{equation}
v(Z_0)= - \frac{ d \ln g(Z_0)}{d Z_0}, 
\label{classical2}
\end{equation}
and, in dimensionless units, the two constraints 
\begin{equation}
\int_0^{ Z_0} \frac{1}{S (Z,Z_0)}dZ=1,
\label{isochrone2}
\end{equation}
and
\begin{equation}
\int_Z^{D} \frac{ g(Z_0)}{ S(Z,Z_0)} dZ_0 = \phi_p(Z),
\label{isochrone}
\end{equation}
apply. This yields an expression for the dimensionless speed 
at position $Z$, $S(Z,Z_0)$ and the normalization condition in equation (\ref{isochrone2}), corresponding to the isochronicity constraint for the effective particle 
rolling down an incline \cite{Matsen3}.
 In the absence of excluded-volume interactions \cite{Zhulina,Zhulina2}, the normalization condition in equation (\ref{isochrone2}) can be written as an 
Abel-Volterra integral equation of the first kind \cite{Tricomi}. Its 
solution leads to a parabolic form for the potential $U(Z)$ \cite{Milner2,Semenov}.
Neglecting the term related to the end-monomer tension, i.e. setting $v(Z_0) \approx 0$, one finds the integral equation
\begin{equation}
\int_{ U}^{ U_0} dt f(t) (t-U)^{-1/2}=1,
\end{equation}
and its solution, in terms of the Riemann-Liouville fractional derivative 
of order $1/2$, as discussed in Appendix \ref{APP:C}, reads
\begin{equation}
f(U)= \sqrt{ \frac{3}{2}} \frac{ dZ}{dU}= 
 - \frac{1}{ \pi} \frac{ d }{ d U} \int_{ U}^{U_0} dt ( t - U)^{-1/2}.
\label{sst}
\end{equation}
The associated potential form is obtained as 
\begin{equation}
U(Z)= \frac{ 3 \pi^2}{8} ( H^2-Z^2),
\label{sst2}
\end{equation}
and the electrostatic potential $\psi^a(Z)=N_c^{-1/2}U(Z)$ follows a 
parabolic form, according to strong-stretching theory \cite{Zhulina}. 
We express distances in dimensionless rescaled units 
$Z'=Z/N_c^{1/2}$, where $H$ represents the brush height in units of $Z$.
The relative charge ratio per chain reads $Q_1=1-\frac{H}{ \gamma}$, where 
\begin{equation}
\gamma= \frac{ 4}{3 \pi^2}L_B N_c^{3/2}. 
\end{equation}
The rescaled counter-ion distribution is given by
\begin{equation}
N_c^{1/2} \phi_c^a(Z)=\frac{3 \pi^2 H^2}{8 \gamma} \exp \{ \psi^a(Z) \},
\end{equation}
and the electrostatic potential outside the brush is written as 
\begin{equation}
\psi^b(Z)=-2 \ln \Big (  \gamma^{1/2}(Z-H+1/H)\Big ),
\end{equation}
where the brush height \cite{Zhulina} is given by 
\begin{equation}
H + \frac{\sqrt{ \pi}}{2} \sqrt{ \frac{3 \pi^2}{8}}H^2 
e^{\frac{3 \pi^2}{8}H^2}  erf (\sqrt{\frac{3 \pi^2}{8}}H)= \gamma,
\end{equation}
the monomer density distribution, 
\begin{equation}
N_c^{1/2}\phi_p(Z)=N_c^{1/2}\phi_c^a(Z)+1 /\gamma,
\label{phip}
\end{equation}
and the counter-ion distribution outside the brush is given by the 
Gouy-Chapman form,
\begin{equation}
 \phi_c^b(Z)=\frac{1}{\gamma}(Z-H+1/H)^{-2}.
\end{equation}
The size of the brush can be expressed as 
\begin{equation}
\frac{\langle Z \rangle }{H} =\frac{H^2}{2 \gamma} e^{  \frac{3 \pi^2}{8} H^2}.
\label{thickness}
\end{equation}
Finally the end-monomer distribution can be obtained, as detailed in the 
Appendix \ref{APP:C}, solving \cite{Zhulina} a similar Abel-Volterra 
equation of the first kind to equation (\ref{isochrone2}), that corresponds to 
the constraint of equation (\ref{isochrone}). One finds, 

\begin{eqnarray}
\label{gz}
&& \gamma N_c^{1/2}g(Z)= \sqrt{ \frac{3 \pi^2}{8}} Z (1+\frac{3  \pi^2H^2}{8}) 
\big ( \psi^a(Z) \big )^{-1/2} \nonumber \\
&& ~~~~+ \sqrt{ \pi} \frac{ 3 \pi^2}{8} H^2 
\exp( \psi^a(Z)) { \it erf} \big ( \sqrt { \psi^a(Z)} \big ). \\ \nonumber
\end{eqnarray}

In the following section we will compare the strong-stretching theory 
predictions with the results obtained within self-consistent field theory, 
where the Poisson-Boltzmann and modified diffusion equations 
(\ref{spe1})-(\ref{spe2}) are solved explicitly by finite difference methods.
We consider both the case of excluded-volume and electrostatic interactions 
being present. It is important to note at this point that typical synthetic 
polymers considered in experimental conditions are relatively insensitive 
to solvent conditions and the long range electrostatic interactions will 
dominate in some cases \cite{Zhulina2}. 

\section{Self-Consistent Field Theory-Results}
\label{SectionV}

In the previous section we discussed the main assumptions and results of 
strong-stretching theory and introduced the fundamental equations for the 
monomer density and electrostatic potential in a system of uniformly charged 
polyelectrolyte chains, tethered to a flat interface, in the absence of 
the external field.
The numerical method, used to solve the mean-field saddle point equations 
discussed within self-consistent field theory, is introduced in this 
section and the results throughout discussed and compared with the analytical 
predictions of strong-stretching theory \cite{Zhulina}.
In order to solve the coupled set of saddle-point equations 
(\ref{spe1})-(\ref{spe2}), we used a real-space, finite difference second-order 
method \cite{Henrici}. The modified diffusion equation (\ref{spe1}) was solved 
using a discrete Crank-Nicholson real-space method, as discussed in detail in 
the past\cite{Matsen3}. 
The full Poisson-Boltzmann equation (\ref{spe2}) was solved using an iterative 
Newton-Raphson method, where the partition function for the small species is 
calculated at each iteration step. 
The algorithm we implemented in order to solve the modified 
diffusion equation (\ref{spe1}) includes the Anderson mixing method \cite{Thompson}. 
We observe our algorithm to converge efficiently in a broad range of 
parameter values. 
It is important to mention that we observe the range of convergence, namely 
the highest number of ions per chain the algorithm can reach for a given 
value of the Bjerrum length and brush thickness parameters, 
to depend in a non trivial way on the Anderson mixing method. Avoiding 
the Anderson mixing method and decreasing the mixing parameter $ \mu$ to 
very small values, typically $ \mu =0.01$, improves the range of convergence 
but increases the time of convergence significantly.
A typical threshold value we obtained and report for comparison: 
$N^{max}_c \simeq 200$, for values of $L=2$ and $L_B = 0.1$ of the 
brush thickness and Bjerrum length parameters respectively. 

\begin{figure}
\includegraphics*[angle=0,scale=0.7]{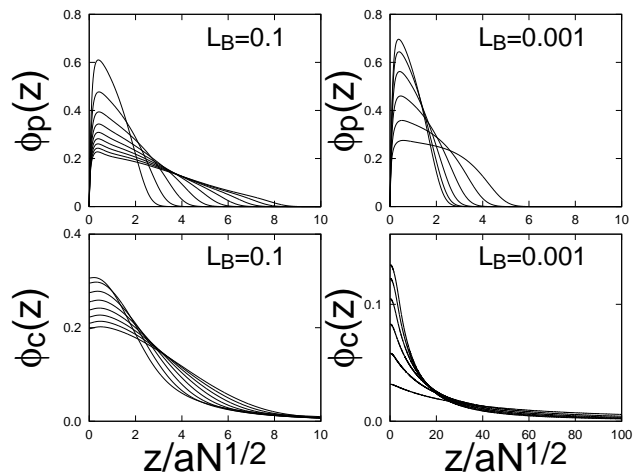}
\caption{ The two upper plots represent the monomer density profile for 
the value of the brush thickness $L/aN^{1/2}=2$ and for increasing values 
of the number of ions per chain $N_c=10 \cdots 90$ and 
$N_c=30,60, \cdots 180$ , for the values of the Bjerrum Length parameter 
$L_B=0.1$ and $ L_B=0.001$ respectively. The two lower plots represent 
the corresponding counter-ion concentration profiles for $N_c=20 \cdots 100$, 
$N_c=30,60, \cdots, 180$.
\label{fig1}}
\end{figure}

A first set of numerical results has been obtained for values of the brush 
thickness parameter $L=2$ and Bjerrum length values $L_B=0.1$ and $L_B=0.003$. 
The results for the monomer density and counter-ion distributions, 
for increasing ionic strength $N_c=10, \cdots , 100$ and  $N_c=30, \cdots 180$ for two different values of the Bjerrum length parameter $L_B=0.1$ and 
$L_B=0.001$ are shown in Fig~\ref{fig1}. Another set of calculations have been 
performed at the values of $L \simeq 1.88$, $L_B \simeq 0.68$ to confirm 
the agreement with closely related calculations \cite{Orland2}. 
We note that, when exceeding the values of the number of ions per 
chain $N_c =30$, considered in that paper as a limiting threshold, 
a peak at low values of the distance appears, already discussed \cite{Taniguchi}.
Our findings resolve the inconsistency reported \cite{Orland2} for 
the monomer density profiles at small values of the grafting distance 
$z/aN^{1/2}~ \simeq 0.05$, where a peak, as can be seen by close inspection 
of Fig.~\ref{fig1} appears. The presence of this peak has already been 
reported \cite{Taniguchi} when considering high values of the ion-per-chain 
parameter $N_c>50$. 

All calculations of this and the following sections were performed 
considering value of the rescaled distance $D=d/aN^{1/2}$ between the grafting surface and 
the electrode to be large enough, so to measure a vanishing counter-ion 
density at the boundaries. 
In order to obtain such condition, at the lowest values of the 
Bjerrum length parameter considered, e.g. $L_B=0.0003$, we needed to consider 
values of order $D=10^3$. 
This was possible as the two grids we consider to solve the modified 
diffusion equation (\ref{spe1}) and equation (\ref{spe2}) did not have the same size, 
the second grid exceeding the size of the first, depending on the 
Bjerrum length values considered. 
This can be seen looking at the range of distances $z/aN^{1/2}$ we plot in Fig.~\ref{fig1} 
for different values of the Bjerrum length parameter $L_B$.

Another method introduced in the past to avoid the numerical self-consistent field 
theory approach is to linearize \cite{Marcelja} the Poisson-Boltzmann equation (\ref{spe2}) 
and to modify the strong-stretching theory accordingly.
As a third, intermediate option, we considered explicitly the Poisson-Boltzmann equation 
(\ref{spe2}) and combined its numerical solution with the analytical form of the 
potential, as predicted by strong-stretching theory.
We hence assume a parabolic form for the effective potential 
$w(Z)$ and solve numerically the Poisson-Boltzmann 
equation (\ref{spe2}), and an estimate for the monomer concentration 
profile, together with the electrostatic potential, follows directly.
The procedure is iterated until convergence is achieved. 
A comparison with the results obtained by the full solution of 
equations (\ref{spe1}) and (\ref{spe2}) is shown in Fig.~\ref{fig3}. 
Closely related methods and results have also been discussed \cite{Misra}.
\begin{figure}
\includegraphics*[angle=0,scale=0.7]{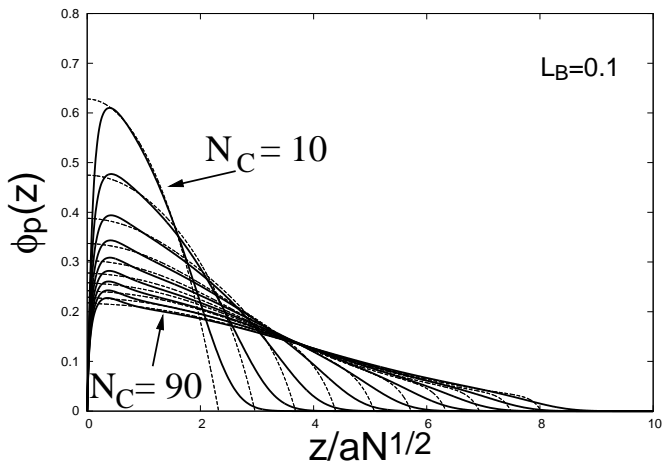}
\caption{Solid lines represent the self-consistent field theory results for 
the monomer concentration profiles for values of the brush thickness parameter 
$L/aN^{1/2}=2$ and Bjerrum length $L_B=0.1$, for increasing values of the ion 
number parameter $N_c=10, \cdot \cdot , 90$. The dashed lines are the results 
obtained with the numerical strong-stretching approach discussed in the text.
  \label{fig2}}
\end{figure}

We now consider the case where excluded-volume interactions are neglected \cite{Zhulina2}. The validity of this assumption \cite{Zhulina} is tested and comparison between polyelectrolyte scaling and strong-stretching theory and the numerical results obtained within self-consistent field theory, when solving the saddle point 
equations (\ref{spe1})-(\ref{spe2}) is presented. We assume here the relative strength of the excluded-volume interactions to be small with respect to the average electrostatic interaction strength, so that the solvent can be considered as a $ \theta$ solvent for uncharged chains. 
This assumption, already discussed and justified in detail \cite{Zhulina2}, 
will be considered in the results of Figs.~\ref{fig3}-~\ref{fig6}, so to compare with scaling and strong-stretching theory.

\begin{figure}
\includegraphics*[angle=0,scale=0.7]{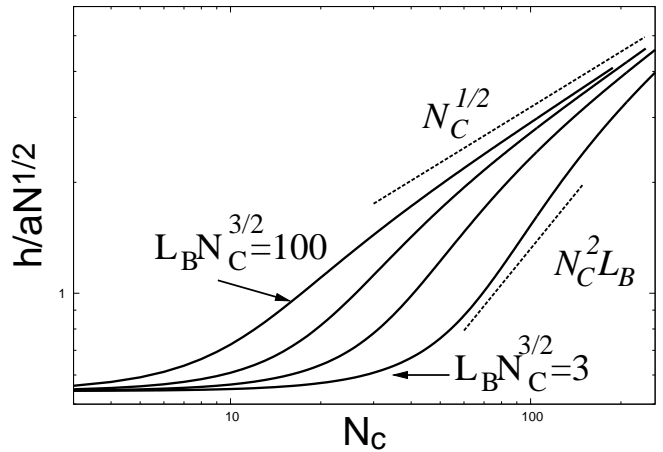}
\caption{ Brush thickness $h/aN^{1/2}$ as a function of the number of ions per chain $N_c$, 
for different values of the Bjerrum length $L_B=0.003,0.01,0.03,0.1$, 
neglecting excluded-volume interactions. 
The two dotted lines are the scaling predictions corresponding to the osmotic 
and Pincus regimes \cite{Pincus} respectively.
\label{fig3}}
\end{figure}

In Fig.~\ref{fig3} we present the results for the brush thickness $h/aN^{1/2}$, as 
obtained from the first moment of the monomer concentration profile, 
for increasing values of the number of ions per chain parameter $N_c$ and for 
vanishing values of $L/aN^{1/2}$.
The classical scaling predictions, discussed in the literature \cite{Pincus}, 
are clearly confirmed, given in particular the large values of the number 
of ions per chain parameter our algorithm was capable to achieve.
The data fit the asymptotic behavior expected at large values of the 
number of ions per chain parameter $N_c$, as shown by the upper dotted line of Fig.~\ref{fig3}, 
where the condition of static equilibration of the osmotic pressure dictates \cite{Pincus}, 
in the strongly charged regime, the scaling law $h/aN^{1/2} = N_c^{1/2}$.
According to the results presented in Fig.~\ref{fig3}, an 
intermediate cross-over regime, the so-called Pincus-regime \cite{Pincus}, is 
present. 
The curves, corresponding to different values of the Bjerrum length parameter 
$L_B=0.003,0.01,0.03,0.1$, cross-over to the intermediate scaling regime, 
according \cite{Pincus} to the scaling expression $h/aN^{1/2} = N_c^{2}L_B$, 
corresponding to the slope of the second dashed line in Fig.~\ref{fig3}.
The region for this crossover to occur is quite narrow: all curves 
in Fig.~\ref{fig3} converge to the neutral brush thickness \cite{Netz} 
value $h_o/aN^{1/2} \simeq 0.54281$, thus explaining why evidence of the 
Pincus regime in numerical simulations  turns out to be an elusive 
task \cite{Netz2}. 
In Fig.~\ref{fig4} we compare the numerical results obtained by 
the numerical self-consistent field theory with the predictions of 
strong-stretching theory, as described by equation (\ref{phip}), 
for the four values of the strong-stretching parameter 
$L_BN_c^{3/2}=3,10,30,100$. 
At large values of $ \gamma$ the agreement between the numerical results and 
the theory becomes very good, except at small values of the 
rescaled distance $Z/N_c^{1/2} < 0.05$, and at the edge of the brush, as expected.

\begin{figure}
\includegraphics*[angle=0,scale=0.7]{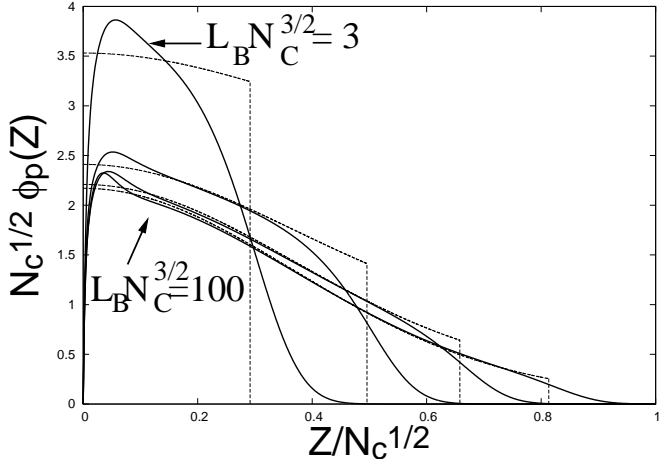}
\caption{ Monomer concentration profile as a function of rescaled distance 
from the grafting surface $Z/aN_c^{1/2}$, for different values of the 
strong-stretching parameter $L_BN_c^{3/2}=3,10,30,100$. Solid lines are 
the results obtained with self-consistent field theory. The dashed 
lines are the predictions of strong-stretching theory. 
Agreement between strong-stretching and the self-consistent field theory 
becomes very good for large values of the parameter $\gamma$.
\label{fig4}}
\end{figure}

\begin{figure}
\includegraphics*[angle=0,scale=0.7]{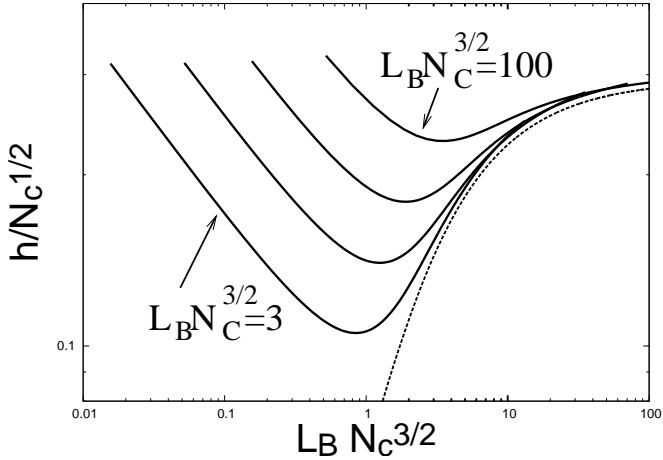}
\caption{ Brush thickness as obtained from the first moment of the monomer 
concentration profile predicted by self-consistent field theory (solid-lines), 
for four different values of the Bjerrum length $L_B=0.0003,0.001,0.003,0.1$ 
and the prediction of strong-stretching theory in equation (\ref{thickness}) as a 
function of the strong-stretching parameter $\gamma$. 
The agreement becomes very good at large values of $\gamma$. 
A small discrepancy though is observed. The reason for this discrepancy
is not clear.
\label{fig5}}
\end{figure}

\begin{figure}
\includegraphics*[angle=0,scale=0.7]{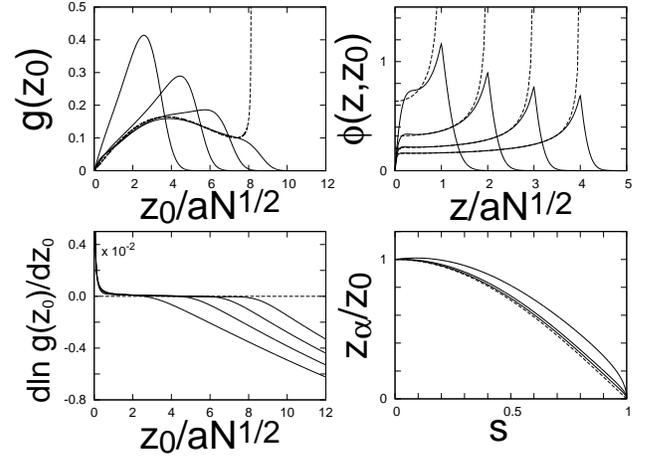}
\caption{ The left upper plot represents the end-segment distribution 
$g(z_0)$. The solid curve corresponds to the results obtained within 
self-consistent field theory, at values of the strong-stretching 
parameter $L_BN_c^{3/2}=3,10,30,100$. The dashed line corresponds 
to the prediction of equation (\ref{gz}) for $L_BN_c^{3/2}=100$. 
The agreement is good for all values of the distance from the grafting 
surface $z/aN^{1/2}$, except at large values where the prediction of 
equation (\ref{gz}) diverges. 
In the lower left plot we show the chain-end tension for the four values 
of the strong-stretching parameter above. 
The upper right plot shows the segment profiles $\phi(z,z_0)$ for 
$L_BN_c^{3/2}=100$ for the four values of the end position 
$z_0/aN^{1/2}=1,2,3,4$. The corresponding prediction of strong-stretching 
theory is represented by the dashed lines. 
The corresponding average polymer trajectories are shown in the lower right 
plot together with the prediction of strong-stretching theory, shown by the 
dashed line.
  \label{fig6}}
\end{figure}

\begin{figure}
\includegraphics*[angle=0,scale=0.7]{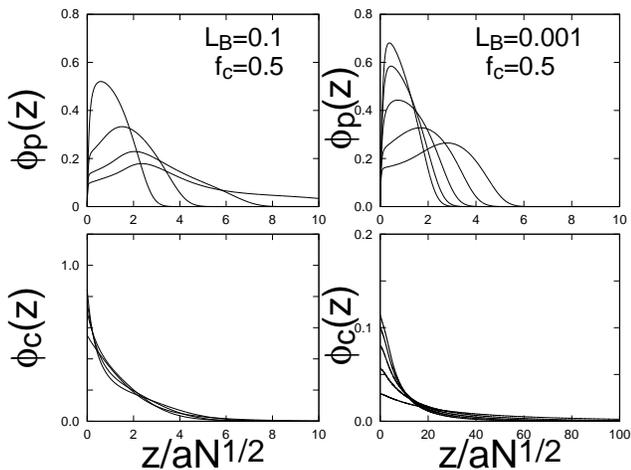}
\caption{ Monomer concentration $\phi_p(z)$ and counter-ion distribution 
profiles $\phi_c(z)$, for brush thickness values of $L/aN^{1/2}=2$ and 
for two values of the Bjerrum length parameter $L_B=0.1$ and $L_B=0.001$, 
as a function of the distance from the grafting surface, for positive vales 
of the relative surface charge parameter $f_e=0.5$, for increasing values 
of the number of ions per chain parameter $N_c=10, \cdots , 40$ and 
$N_c=30,60, \cdots,150$ respectively.
\label{fig7}}
\end{figure}

\begin{figure}
\includegraphics*[angle=0,scale=0.7]{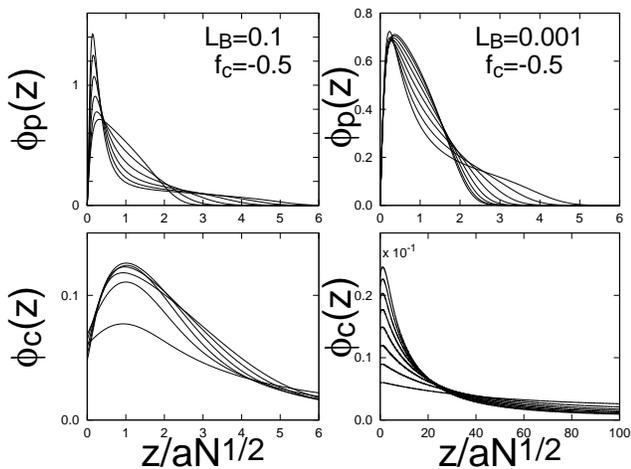}
\caption{Monomer concentration $\phi_p(z)$ and counter-ion distribution 
profiles $\phi_c(z)$, for brush thickness values of $L/aN^{1/2}=2$ and 
for two values of the Bjerrum length parameter 
$L_B=0.1$ and $L_B=0.001$, as a function of the distance from the 
grafting surface, for negative vales of the relative surface charge 
parameter $f_e=-0.5$ and for increasing values of the number of ions per 
chain parameter $N_c=10, \cdots , 60$ and $N_c=30,60, \cdots,240$ respectively.
\label{fig8}}
\end{figure}

A few conclusive remarks, while closing this section, are here in order.
As shown in Fig.~\ref{fig6}, the fundamental assumption of strong-stretching 
theory becomes more accurate for increasing values of the 
strong-stretching parameter $L_B N_c^{3/2}$.
At values of $L_BN_c^{3/2}=100$, the chain-end tension, as measured within 
self-consistent field theory, approaches zero, for all values of the 
distance $z/aN^{1/2}$ inside the brush. Fig.~\ref{fig6} also 
shows the segment profiles $\phi(z,z_0)$, for chains with different 
end-positions $z_0$ at values of $L_BN_c^{3/2}=100$ and the corresponding 
average polymer trajectories $z_{\alpha}(s)/z_o$ as obtained within 
self-consistent field theory, together with the prediction of 
strong-stretching theory.
We finally present the results obtained within self-consistent field 
theory, when considering a uniform charge on the grafting surface. We  
obtained results for both the case of similarly and oppositely charged surface. 
The monomer-density and counter-ion distribution profiles are shown in 
Fig.~\ref{fig7} and Fig.~\ref{fig8}. 

The results were obtained for both positive and negative 
values of the relative surface charge parameter $f_e$, for a brush thickness 
parameter value $L/aN^{1/2}=2$. The boundary condition in equation (\ref{bc2}) has been 
changed accordingly. As expected, for the case 
of a similarly charged surface, the polymer brush swells, and the monomer-density 
profile obtained at small values of the relative surface charge parameter changes its 
shape and deviates from the parabolic form. Similarly, when considering an oppositely 
charged grafting surface, as in Fig.~\ref{fig8}, the brush contracts and the counter-ion 
distribution changes self-consistently. Related results, for the case of 
a similarly charged grafting surface, have been reported in the 
past \cite{Zhulina2}. 

\section{Charged-end-group polymer brushes}
\label{SectionVI}

In the previous section we presented numerical self-consistent field theory 
results for uniformly charged, end-grafted polyelectrolyte systems, 
both in the presence and in the absence of excluded-volume interactions and 
discussed the effects of a uniform electric field. We discussed the results for the chain-end, 
counter-ion and monomer-density distributions, for different values of the 
relative surface charge parameter $f_e$. The analysis suggested that the 
parabolic form of the monomer density profile is drastically affected by the 
electric field, both in the similarly and the oppositely charged case. 

\begin{figure}
\includegraphics*[angle=0,scale=0.7]{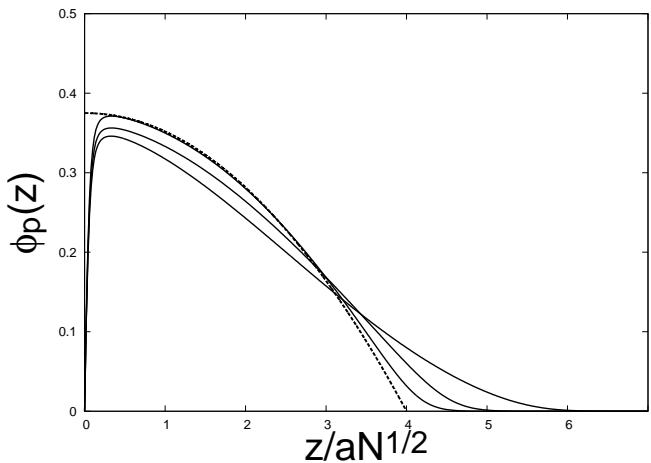}
\caption{Monomer density profiles for a brush with thickness parameter $L/aN^{1/2}=4$, Bjerrum length parameter value 
$L_B=0.1$ for increasing values of the number of ion-per-chain-end parameter $N_c=1,10,20$. The dashed line 
represents the prediction of strong-stretching theory, in the absence of charged-end-groups.
\label{fig9}}
\end{figure}

We consider in this section the case of a polymer brush in solution, 
characterized by charged end-groups \cite{Zhulina3,Andelman} in the presence of counter-ions 
and excluded-volume interactions between the monomers. 
Rather than considering a uniformly charged polyelectrolyte, as discussed in section \ref{SectionV}, we study in this section the case of grafted neutral chains, with a functional charged end-group, represented in our model 
by the terminal, non-grafted $\chi N$ monomers along the chain. As discussed above, we consider the case 
of strong polyelectrolytic functional groups, so that the charges are assumed to stay bounded to the chain-ends. 
As for the case of uniformly charged polyelectrolytes, the $N_c$ ions per chain
are uniformly smeared along the terminal, non-grafted section of the chain, with a rescaled charge density $f/ \chi$ 
so that, consistently with the notation of Appendix \ref{APP:A}, we consider the charge 
distribution along the chains to be given by $z_p(s)=f/ \chi \Theta( s- \chi)$. 
We solve the saddle-point equations (\ref{spe1})-(\ref{spe2}) that now 
involve both the charged and uncharged monomer densities and the 
effective field $w_p(z)$ is also defined accordingly. 
In Fig.~\ref{fig9} we show the total (charged and uncharged) monomer density distribution of a brush 
of thickness $L/aN^{1/2}=4$, for increasing values of the number of ions per chain-end parameter $N_c=1,10,20$. 
For small values of the ions per chain parameter, namely  $N_c=1$, the electrostatic interactions 
between the chain-ends are negligible, as can be seen in Fig.~\ref{fig9} and the monomer distribution 
is close to the parabolic prediction of strong-stretching theory, the brush thickness parameter $L/aN^{1/2}$ 
being relatively large, as expected.
Deviations from the parabolic profiles can be observed and are due to the inter-chain-end electrostatic 
interactions at higher values of the ion-per-chain-end parameter $N_c$. 
The calculation has been performed for the specific value of $ \chi=1/30$, so that the size of the 
polymer chain is about thirty times the size of the chain-end group. We performed several test runs and 
checked that for smaller sizes of the chain-end group the monomer density profiles did not change significantly.
For increasing values of $N_c$ the brush swells, due to the repulsive electrostatic chain-end interactions 
and the counter-ion distribution changes accordingly.
We now discuss the response of the system to an electric field. In our analysis we consider small values of the number of ions per chain-end parameter, namely $N_c=1$, so to minimize the interaction between the chain-ends.

As discussed in section \ref{SectionIV}, we consider the brush to be grafted to a flat surface that acts 
as an electrode, while the second electrode is kept at a distance $D \gg L$. In order to study the response of the 
system to the uniform electric field, we solved the saddle point equations (\ref{spe1}) and (\ref{spe2}) changing 
the boundary conditions for the electrostatic potential as well as modifying the effective field accordingly. Fig.~\ref{fig11} shows the monomer density 
profiles for the brush of thickness $L/aN^{1/2}=4$, Bjerrum length $L_B=0.1$ and number of ions per 
chain-end parameter $N_c=1$ for both positive and negative values of the surface charge $f_eL_B= \pm 2$. 
The neutral surface charge case $f_e=0$ is also shown. As expected, we observe deviations from the parabolic 
profile for finite values of the electric field strength. Differently, for vanishing values of the surface charge 
$f_e=0$, the profile is very close to the strong-stretching parabolic prediction, the value of the brush thickness 
$L/aN^{1/2}=4$ being relatively large \cite{Matsen3}. In the following section we discuss how to generalize strong-stretching theory for the case of a uniform electric field, assuming the relative charge at the 
chain-ends to be small. 

\begin{figure}
\includegraphics*[angle=0,scale=0.7]{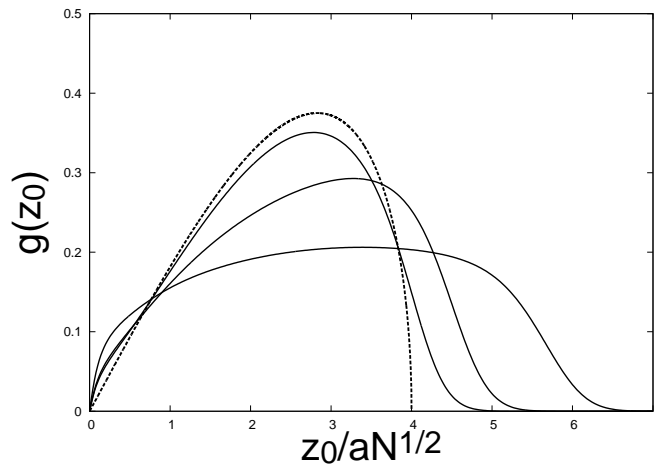}
\caption{ Chain-end distribution $g(z_0)$ for a brush with thickness parameter $L/aN^{1/2}=4$, Bjerrum length 
parameter value $L_B=0.1$ for increasing values of the number of ion-per-chain-end parameter $N_c=1,10,20$. 
The dashed line represents the prediction of strong-stretching theory, in the absence of charged-end-groups.
\label{fig10}}
\end{figure}

\section{Strong-stretching theory of charge-end-group polymer brushes}
\label{SectionVII}

The mathematical foundations of strong-stretching theory stem from the 
fact that the isochronicity constraint in equation 
(\ref{isochrone2}) may be regarded as an Abel-Volterra integral equation 
of the first kind.
This determines uniquely the form of the potential 
\cite{Milner2}. To determine how a uniform tension at the chain-ends,  
induced by the electric field, affects the potential form $U(Z)$, we 
evaluated the corresponding corrections, solving the isochronicity constraint 
condition in equation (\ref{isochrone2}). The presence of the electric field 
induces a chain-end uniform tension and converts the associated 
integral equation from an Abel-Volterra equation of the first kind to an 
Abel-Volterra equation of the second kind, as also discussed in detail in 
Appendix \ref{APP:C}. This leads to evaluate the main corrections to the 
parabolic potential, due to the presence of the field, as we now derive.
Let us consider directly the constraint in equation (\ref{isochrone2}),  
that can be written, following the few steps in Appendix \ref{APP:C},
\begin{equation}
\int_{ \eta}^{U_0} dt (t- \eta)^{-1/2}= \int_{ \eta}^{U_0} dt f(t) 
 \Big \{ \pi- \alpha (t-\eta)^{-1/2} \Big \},
\label{SST2a}
\end{equation}
where $f(U)= \sqrt{\frac{3}{2}}\frac{dZ}{dU}$ is related to the first 
derivative of the inverse function $Z(U)$.
We then find corrections to the strong-stretching results. 
Namely, after a few steps, we obtain the following Abel-Volterra integral 
equation of the second kind:
\begin{eqnarray}
\frac{df( \eta)}{d \eta} &-& \frac{ \pi}{\alpha^2} f( \eta) 
=  \frac{ d}{ d \eta} F( \eta)  \nonumber \\
F( \eta)= g( \eta)&-& \frac{1}{ \alpha} \int_{ \eta}^{U_0} dt g( \eta) 
(t- \eta)^{-1/2}, \\ \nonumber 
\end{eqnarray}
where $g( \eta)=-1/ \alpha$, and where $ \alpha = \sqrt{6} V_0$ measures 
the uniform end-monomer force induced by the electric field at the chain ends.
The solution to this Abel-Volterra integral equation of the second kind 
can be obtained as follows:
\begin{equation}
f( \eta)= \frac{ dZ}{ d \eta}= \frac{1}{ \alpha} e^{ \frac{\pi}{ \alpha^2}(U_0- \eta)} 
\frac{1}{ \sqrt{\pi}} \Gamma \Big( \frac{1}{2},\frac{\pi}{ \alpha^2}
(U_0- \eta) \Big),
\end{equation}
$\Gamma( \alpha,x)$ being the incomplete gamma function of order $\alpha$.
The above expression can be integrated and the corresponding equation
for the potential reads 

\begin{equation}
\sqrt{\frac{3}{2}} Z= \frac{2}{\pi} (U_0-U)^{1/2}+\alpha- \frac{ \alpha}{\sqrt{\pi}} e^{ \frac{\pi}{ \alpha^2}(U_0- \eta)} 
\Gamma \Big( \frac{1}{2},\frac{\pi}{ \alpha^2}
(U_0- \eta) \Big).
\label{SST2p}
\end{equation}
The expression above reduces to the parabolic strong-stretching potential 
when the chain-end tension vanishes, as expected. 
From the potential in equation (\ref{SST2p}) , we determined the monomer-density and 
chain-end distributions by means of the field equation (\ref{field}), in the 
limit of weakly charged end-groups. This assumption has been justified and 
discussed in Fig.~\ref{fig10}.
A similar analysis, detailed in Appendix \ref{APP:C}, leads to an explicit 
expression for the chain-end distribution $g(Z_0)$, for small values of the 
electric field strength $ \alpha$. We find, after a few steps detailed in 
Appendix \ref{APP:C},
\begin{equation}
g( \eta) \frac{ dZ}{ d \eta} = - \frac{2}{ \pi} \eta^{ 1/2} + \frac{ \alpha}{2}- \frac{ \alpha}{ \pi^{3/2}} e^{ \frac{ \pi}{ \alpha^2} \eta} \Gamma \Big( \frac{1}{2},\frac{\pi}{ \alpha^2} \eta \Big).
\label{SST2g}
\end{equation}
Equation (\ref{SST2g}) reduces to the well known expression for the chain-end distribution \cite{Milner2}, when no tension 
is present at the chain-ends.
The value of the chain-end prefactor $ \delta \simeq 0.263$ has been obtained comparing the results of numerical self-consistent field theory, as in the inset of Fig. \ref{fig12}, and the results obtained from equation (\ref{SST2p}) for the brush height $h/aN^{1/2}$. It would be interesting to see if the prefactor can be computed explicitly, and if it depends on the brush thickness parameter $L/aN^{1/2}$.

\begin{figure}
\includegraphics*[angle=0,scale=0.7]{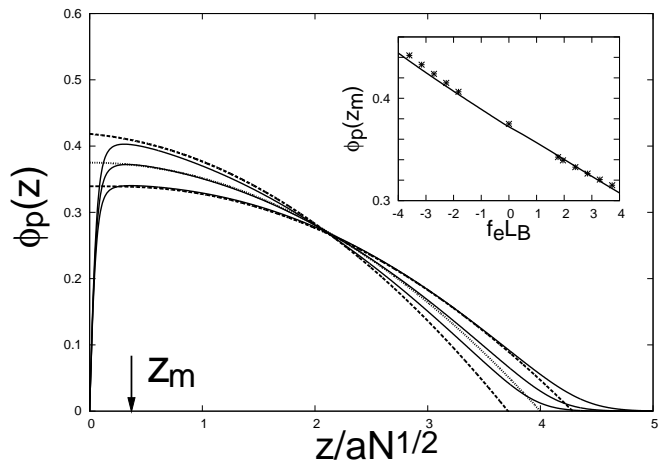}
\caption{Monomer density profile $ \phi_p(z)$ for a brush with thickness parameter $L/aN^{1/2}=4$, Bjerrum length 
parameter value $L_B=0.1$, for values of the number of ion-per-chain-end parameter $N_c=1$, for neutral surface 
charge $f_e=0$, as well as similarly and oppositely charged grafting surface,with rescaled surface charge parameter values 
$f_eL_B= \pm 2.0$. The dotted line represents the prediction of strong-stretching theory, in the absence of charged-end-groups.
The dashed lines represent the prediction of the modified strong-stretching theory (\ref{SST2p}), for values of the parameter $\alpha =\delta f_eL_B$. The inset shows the monomer-density at the reference distance $z_m=0.378$, as measured within 
self-consistent field theory, shown by the solid line, and as predicted by equation (\ref{SST2p}), for different values of the chain-end tension parameter $ \alpha$, shown by the star symbols.
\label{fig11}}
\end{figure}

Equations (\ref{SST2p}) and (\ref{SST2g}) generalize the MWC strong-stretching theory to problem of charge-end functionalized brushes in the presence of a 
small, uniform external electric field. A comparison with the results 
of numerical self-consistent field theory is shown in Fig.~\ref{fig11} 
and Fig.~\ref{fig12}.

\begin{figure}
\includegraphics*[angle=0,scale=0.7]{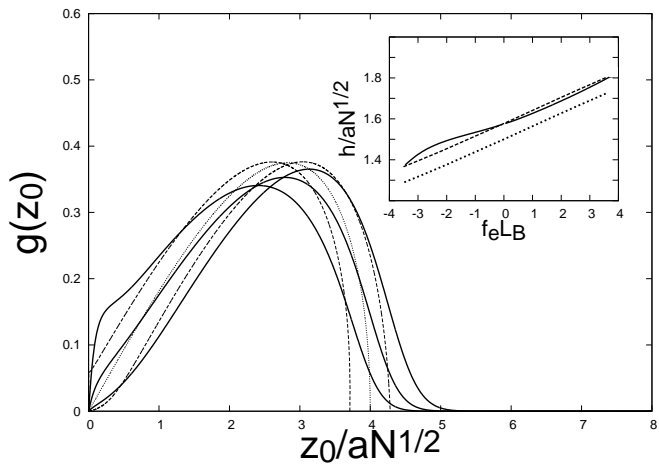}
\caption{Chain-end distribution $ g(z_0)$ for a brush with thickness parameter $L/aN^{1/2}=4$, Bjerrum length 
parameter value $L_B=0.1$ for values of the number of ion-per-chain-end parameter $N_c=1$, for neutral surface 
charge $f_e=0$, as well as similarly and oppositely charged grafting surface,with rescaled surface charge parameter values 
$f_eL_B= \pm 2.0$. The dotted line represents the prediction of strong-stretching theory, in the absence of charged-end-groups.
The dashed lines represent the prediction of the modified strong-stretching theory in equation (\ref{SST2g}), for different values of the chain-end tension parameter $\alpha= \delta f_e L_B$.
The inset shows the size of the brush $h/aN^{1/2}$, as predicted from numerical self-consistent field theory, shown by the solid line, and as obtained according to the modified strong-stretching theory in equation (\ref{SST2p}), shown by the dotted line. The dashed line shows the results of the modified strong-stretching theory, shifted by the zero-field value, obtained subtracting the brush size as obtained from self-consistent field theory and the MWC theory prediction.
\label{fig12}}
\end{figure}

We conclude observing that the monomer-density profile is affected by the uniform electric field in a different way that what seen for uniformly charged polyelectrolyte brushes. Note however that, when comparing the results of this section with the monomer-density profiles of Fig.~\ref{fig7} and Fig.~\ref{fig8} one should keep in mind that the results discussed for the polyelectrolyte problem were obtained at relatively large values of the number of ions per chain parameter $N_c$, the electrostatic interactions between monomers being relatively strong. 
It is interesting to observe that for a uniformly charged polyelectrolyte system,at low values of the number of ions per chain parameter, e.g. $N_c=1$, the effect of the uniform electric field is very similar to what observed for charged-end-group chains and the monomer-depletion effect close to the surface, 
seen in the analysis of Fig.~\ref{fig7} for a similarly charged surface, is not present, as in the case of grafted charge-end-group polymer chains discussed in this section. This suggests that the strong-stretching theory corrections discussed in this work might apply reasonably well for the problem of weakly charged 
polyelectrolyte chains in a uniform electric field. 

\section{Conclusions}
\label{SectionVIII}
In this work we studied the properties of charged polymers grafted to a flat 
interface, in the presence of an external electrical field. Different charge 
configurations have been discussed. We firstly discussed the properties of 
polyelectrolyte brushes, where the charges are uniformly distributed 
along the chains, while counter-ions are in solution and free to diffuse 
between the grafting surface and the second, oppositely charged electrode.
The monomer-density and counter-ion distributions  
have been obtained by numerical self-consistent field theory and comparison 
has been made with the analytical strong-stretching theory predictions. 
The effect of a uniform electric field, both in the oppositely 
and similarly charged case, has been studied. We considered 
charge-end-functionalized, grafted polymer chain systems. As for the case 
of polyelectrolyte brushes, we studied the conformational and electrostatic 
properties of the brush by means of numerical self-consistent field theory. 
In the case of weakly charged, end-group-functionalized polymers we have 
been able to generalize the predictions of strong-stretching theory to the 
case of a uniform electric field. 
We have shown how the corrections to the MWC theory for the monomer-density and chain-end 
distributions can be derived in closed form. The presence of an external field 
converts the corresponding Abel-Volterra equation from an integral equation of the 
first kind to an integral equation of the second kind. Deviations from the parabolic 
profile have been obtained. A comparison of the theory with numerical self-consistent 
field theory has been performed and agreement has been observed. 

\begin{acknowledgments}
This research was founded by EPSRC under grant number EP/F068425/1. 
I acknowledge very useful discussions with Mark Matsen who participated to 
the early stages of this work and Jaeup Kim for providing the algorithm 
that implements the self-consistent field calculation for neutral polymer 
brushes.
\end{acknowledgments}

\bibliography{peb}

\appendix

\section{ \label{APP:A} Partition function and Statistical field theory}
\renewcommand{\theequation}{{}\thesection\arabic{equation}}
\setcounter{equation}{0}

In this appendix we review the basic formalism for a multicomponent 
polyelectrolyte system in solution. The saddle point equations 
(\ref{spe1}) and (\ref{spe2}) for a system of a single species of 
polyelectrolyte chains in the presence of counter-ions, discussed 
and studied in this paper are derived in detail in this Appendix 
and some simplifications, compared to a closely related derivation 
\cite{Shi}, are introduced. 
The average canonical partition function of an inhomogeneous multicomponent 
polyelectrolyte system is written \cite{Fredrickson,Shi} as a functional 
integral over the cartesian position of the $n_k$ small species, namely 
solvent molecules and counter-ions, over the conformation of charged 
polymer chains, belonging to $n_p$ polymer distinct species,  
and finally over the charge distribution in the system, given the 
dimensionless polymer and small species concentrations in equations  
(\ref{density1}) and (\ref{density2}):
\begin{equation}
\langle { \cal Z} \rangle = \prod_p \Big [ \prod_{i=1}^{n_p} 
\sum_{ \{ z_p^i(s) \} } P(\{ z_{p}^i(s) \}) \Big ] { \cal Z}( \{ z_p^i(s) \}),
\end{equation}
where
\begin{eqnarray}
\label{PF}
&&~~~{ \cal Z} = \prod_j \Big ( \frac{ \zeta_j^{n_j}}{n_j!} \Big )
\int \prod_p \big ( \prod_i^{n_p} { { \cal D}} { \bf r}_p^i \Big ) 
\int \prod_k \big ( \prod_{i}^{n_k} d { \bf r}_k^i \big ) \nonumber \\
&& \times \exp \big ( - \frac{ \beta}{2} \rho_0^2 \sum_{ jj'} \int d { \bf r} 
d { \bf r'} \hat{\phi}_j( { \bf r}) { \cal V}_{jj'}( { \bf r} - { \bf r}') 
\hat{\phi}_{j'}( { \bf r}') \big ) \nonumber \\ 
&& ~~~ \times  \delta \Big ( \sum_j z_j N_j n_j \Big )   \prod_j \delta 
\Big ( \int d { \bf r} \hat{ \phi}_j( { \bf r}) - n_jN_j \Big ) \nonumber \\
&& ~~~~~~~~~~~~\times \prod_{ { \bf r}} \delta \Big ( \sum_j  \nu_j 
\hat{ \phi}_j( { \bf r}) - 1 \Big ). \\ \nonumber 
\end{eqnarray}
In the expression above, the chain conformation path integral measure 
$  { \cal D} { \bf r}^p= d { \bf r}^p P_{N_p}( { \bf r}^p)$,  
\begin{equation}
P_{N_p}( { \bf r}_p^i) = \exp \Big ( - \frac{3}{2 a^2 } \int_0^{N_p} ds 
[ \dot{ { \bf r}}_p^i(s)]^2 \Big ),
\end{equation}
is the Wiener measure, where $N_p$ indicates the length of each polymer 
species, $a$ is the Kuhn segment, 
$\zeta_j=\exp^{-\beta \rho_o^2 W_{jj}(0)}/ \lambda_j^3$ is the reference 
chemical potential for species $j$, that includes the contact interaction 
energy and depends \cite{Shi} on the thermal wave-length 
$\lambda_{Tj}=h/ \sqrt{2 \pi m_j k_B T}$.
The canonical partition function has to be averaged over the charge 
distribution $P(\{ z_{p}^i(s) \})$, characterizing each polyelectrolyte 
present in solution independently. The interaction energy term includes both 
short-ranged contact and long-ranged electrostatic interactions,

\begin{eqnarray}
&& { \cal V}_{ij}( { \bf r}- { \bf r}') = { \cal W}_{ij}( { \bf r}- { \bf r}') 
+{ \cal E}_{ij}( { \bf r}- { \bf r}') \nonumber \\
&& ~~~~~~= W_{ij} \delta(  { \bf r}- { \bf r}') + 
\frac{e^2 z_i z_j } { \varepsilon({ \bf r}) | { \bf r}- { \bf r}'|},\\ \nonumber
\end{eqnarray}
where $z_j$ is the valence of each species and should be considered as a 
function of $s$ for any $j$ that corresponds to a polymer species, where 
$\nu_j=\rho_o/ \rho_j^o$ measures the different size of each molecule with 
respect to the reference density $ \rho_o$, $W_{ij}$ is the contact energy 
between species $i$ and $j$, $e$ is the elementary charge unit, 
$\varepsilon({ \vec r})$ the dielectric function and $\beta=1/k_BT$ is the 
inverse temperature in Boltzmann units. The first two constraints in the 
above expression for the partition function relates to charge and particle 
conservation, while the last constraint refers to the incompressibility 
condition. We introduce the set of fields 
$ \omega_j( { \bf r})$, $ \phi_j ( \bf{r})$ and express the canonical 
partition function in terms of a statistical 
field theory \cite{Parisi}, using the particle-to-field transformation,

\begin{eqnarray}
&&\langle { \cal Z} \rangle =
\int \prod_j [ { \cal D} \omega_j { \cal D} \phi_j ] 
\exp \Big [ - \beta { \cal H} \big ( \{ \omega_j \},\{ \phi_j \} \big ) \Big ]  \nonumber \\
&&~~~~~~ \times \prod_{ { \bf r}} \delta \Big (  \sum_j \nu_j \phi_j( { \bf r})-1  \Big ) 
\delta \Big ( \sum_j \bar{z}_j N_j n_j \Big ), \\ \nonumber
\end{eqnarray}
where Stirling's formula has been used and where the canonical partition function is expressed in terms of the single species partition functions ${ \cal Q}_j$, as an integral with respect to density and auxiliary fields of an effective Hamiltonian or 'action', 

\begin{eqnarray}
\label{H}
&&~~~{ \beta \cal H} \big ( \{ \omega_j \}, \{ \phi_j \} \big ) 
= - i \rho_0 \sum_j \int d { \bf r} \omega_j( { \bf r}) \phi_j( { \bf r}) 
\nonumber \\ 
&&~~~~~~~ + \frac{ \beta \rho_0^2}{2} \sum_{jj'} \int d{ \bf r} d { \bf r}' 
\phi_j( { \bf r}) { \cal V}_{jj'}  \phi_{j'}( { \bf r}')
\nonumber \\
&&-  \rho_0 \sum_j \frac{ \bar{ \phi}_j V}{N_j} \ln \frac{ { \cal Q}_j}{ \bar{ \phi}_j}  +  \rho_0 \sum_j \frac{ \bar{ \phi}_j V}{N_j} 
\ln \frac{ { \zeta}_jN_j}{\rho_0} \\ \nonumber
\end{eqnarray}
where $ \bar{ \phi}_j = N_jn_j/ \rho_o V $ enforces the particle conservation 
constraint, where  $V$ is the volume, where the single species partition 
functions are:

\begin{eqnarray}
{ \cal Q}_k &=& \frac{1}{V} \int d { \bf r} 
\exp \Big \{ -i \int d { \bf r}^k   \omega_k( { \bf r}) \Big \} \nonumber \\
{ \cal Q}_p&=& \frac{1}{V}  \int { \cal D} { \bf r}^p(s)  
\sum_{ \{ z_p(s) \} } P(\{ z_p(s)\}) \nonumber \\
 &\times& \exp \Big \{ -i \int_0^{N_p} ds 
[  \omega_p( { \bf r}_p(s)) \Big \}, \\ \nonumber
\end{eqnarray}
and where finally we have considered the same charge distribution within 
each species being 

\begin{equation}
\bar{z}_p=\int_0^1ds z_p(s) P(z_p) = z_p f,~~\bar{z}_k=z_k.
\end{equation}
Both the smeared and the annealed charge distribution case are included in 
the present formulation, while the quenched distribution problem 
requires a different formulation. The saddle point equations 
corresponding to equation (\ref{H}), up to a redefinition of the fields $\omega_j$ are,

\begin{eqnarray}
\label{SP1}
&& ~~~~ - i \omega_i - \frac{ \beta \rho_o}{2} 
\sum_j \int d { \bf r}' { \cal V}_{ij}( { \bf r} - { \bf r}') \phi_i({ \bf r}')
  d{ \bf r}'+ \eta ({ \bf r}) \nu_i=0 \nonumber \\  
&& ~~~~ - i \phi_i= \frac{ \bar{ \phi_i}V}{N_j { \cal Q}_j} 
\frac{ \delta { \cal Q}_j}{ \delta \omega_j} \\ \nonumber
\end{eqnarray}
where $ \eta( { \bf r} ) $ is the Lagrange multiplier corresponding to the 
incompressibility constraint and 
where charge conservation applies: $ \sum_j \bar{z}_jN_jn_j=0 $.
The second set of saddle point equations, for the polymeric and small 
species, explicitly reads:
\begin{eqnarray}
\label{SP2}
\phi_p&=&\frac{ \bar{ \phi}_p}{{ \cal Q}_p} 
\int_0^1 ds q_p( { \bf r},s)q_p^{ \dagger}({ \bf r},s) \nonumber \\
\phi_k&=&\frac{ \bar{ \phi}_k}{{ \cal Q}_k} 
\exp \{ -i \omega_k( { \bf r}) \} \\ \nonumber 
\end{eqnarray}
where the polymer propagator $q_p({ \bf r},s)$ is the solution of the 
modified diffusion equation, 
\begin{eqnarray}
\label{propagator2}
 \frac{ \partial}{ \partial s} q_p({ \bf r},{ \bf r}_0,s) &=& \frac{a^2N_p}{6} 
\nabla^2q_p( { \bf r},{ \bf r}_0,s)
-i \omega_p({ \bf r},s) q_p({ \bf r},{ \bf r}_0,s), \nonumber \\
 q_p({ \bf r},{\bf r}_0,1)&=& (a^2N)^{3/2} \delta ({ \bf r}-{ \bf r}_0). \\ \nonumber
\end{eqnarray}
\subsection*{Integration of the Solvent Degrees of Freedom}
It is useful to rewrite the effective hamiltonian in equation (\ref{H}) and the 
corresponding saddle point equations for the case where a single solvent 
species is present \cite{Shi}. One finds,

\begin{eqnarray}
&&{ \beta \cal H} \big ( \{ \omega_j \}, \{ \phi_j \} \big ) 
= - i \rho_0 \sum_j \int d { \bf r} \omega_j( { \bf r}) \phi_j( { \bf r}) \nonumber \\ 
&& ~~~~~~~~~~+ \rho_0 \sum_{jj'} \int d{ \bf r} \chi_{jj'}  \phi_j( { \bf r}) 
\phi_{j'}( { \bf r}) \nonumber \\ 
&& ~~~~~~+ \frac{ \beta \rho_0^2}{2} \sum_{jj'} \int d{ \bf r} d { \bf r}' 
\phi_j( { \bf r}) { \cal E}_{jj'}  \phi_{j'}( { \bf r}') \nonumber \\
&& ~~~-  \rho_0 \sum_j \frac{ \bar{ \phi}_j V}{N_j} \ln \frac{ { \cal Q}_j}{ \bar{ \phi}_j} 
+  \rho_0 \sum_j \mu_{0j} \bar{\phi}_j, \\ \nonumber
\end{eqnarray}
where,
\begin{eqnarray}
\label{chi}
\mu_{0j}&=&\frac{ \beta \rho_0}{2} \Big [ W_{jj}-\frac{1}{N_J} 
\ln \Big ( \frac{ \zeta_j Z_j}{ \rho_0} \Big ) \Big ] \nonumber \\
\chi_{ij}&=& \frac{ \beta \rho_o}{2} \Big [  2 W_{ij}- 
 ( \frac{ \nu_j}{ \nu_i}W_{ii}+ \frac{ \nu_i}{ \nu_j}W_{jj}) \Big ].\\ \nonumber 
\end{eqnarray}

The saddle point equations for the solvent degrees of freedom can be 
integrated out. In the semi-dilute regime \cite{Shi}, the saddle point 
equation for the partition function ${ \cal Q}_s$ defined in equation (\ref{SP2}) 
can be expanded to obtain, 
\begin{equation}  
i \omega_s \simeq \sum_{j \ne s}  \nu_j \phi_j,
\end{equation}  
where the primed sum is over all species but the solvent species. 
The Lagrange multiplier $ \eta( { \bf r})$ can be written as 
\begin{equation}  
 \eta( { \bf r} ) \nu_s= \sum_{ j \ne s} \Big ( \nu_j- \chi_{js} \Big ) 
\phi_j({ \bf r})
\end{equation}  
so that a new set of reduced saddle point equations, with $i \ne s$, holds
\begin{eqnarray}
&& ~~ -i \omega_i + \sum_{ j \ne s} v_{ij} \phi_j( { \bf r}) 
+ \frac{ \beta \rho_o}{2} \sum_j \int d{ \bf r}' 
{ \cal E}_{ij}( { \bf r}'-{ \bf r}) \phi_j( { \bf r})\nonumber \\
&& ~~ -i \phi_i= \frac{ \bar{ \phi_i}V}{N_j { \cal Q}_j} 
\frac{ \delta { \cal Q}_j}{ \delta \omega_j}, \\ \nonumber 
\end{eqnarray}
where $v_{ij}= (\frac{ \nu_i \nu_j}{ \nu_s}- \chi_{ij}) $,
where the index $j$ is intended to run over all species excluding the 
solvent and where finally the Flory parameter between species $i$ and $j$ 
is defined in equation (\ref{chi}).

\section{\label{APP:B} Derivation of Model Details}
\renewcommand{\theequation}{{}\thesection\arabic{equation}}
\setcounter{equation}{0}

We consider explicitly the case where we can neglect the excluded-volume 
interactions of the small species. For two distinct indices for the polymeric 
and small species, considering the dielectric constant to be uniform across 
the system and considering the case of a smeared polyelectrolyte charged 
distribution, with charged ion fraction parameter $f$, we find:

\begin{eqnarray}
 i \omega_k( { \bf r}) &=& \rho_0 \frac{ \beta e^2}{2} z_c  \int d { \bf r} 
\frac{ \sum_j z_j \phi_j( { \bf r}')}{ \varepsilon | { \bf r}- { \bf r}'|}, 
\nonumber \\
 i \omega_p( { \bf r}) &=& \sum_{p'} v_{pp'} \phi_{p'}( { \bf r}) + 
i \frac{z_p}{z_k}f \omega_k\\ \nonumber 
\end{eqnarray}
where $i \omega_c$ is related to the dimensionless electrostatic potential 
$\psi( { \bf r})$ according to
$\psi ({ \bf r})=i \omega_k ( { \bf r}) z_k$. Let us consider the case of a 
single polymer species, and let us consider a 
single species of point particles: in this case the above saddle point 
equations reduce to
\begin{eqnarray}
\label{eff}
 i \nabla^2 \omega_c( { \bf r}) &=& - 4 \pi l_B \rho_0 z_c \sum_j z_j 
\phi_j( { \bf r}) \nonumber \\
 ~~~i \omega_p( { \bf r})~~~ &=& v \phi_{p}( { \bf r}) 
+ i \frac{z_p}{z_k}f  \omega_c({ \bf r}). \\ \nonumber 
\end{eqnarray} 
The corresponding mean-field free-energy can be computed according to 
equation (\ref{H}). The canonical Model F field theory partition function 
\cite{Fredrickson} reads:
\begin{equation}
\langle { \cal Z} \rangle = { \cal Z}_0 \int D \omega_p D \omega_c 
\exp \big  \{ - \beta { \cal H} [\omega_p,\omega_c ] \big \}
\end{equation}
where 
\begin{eqnarray}
&&~~~~~~~\beta { \cal H} [\omega_p,\omega_c ]/ \rho_0= \frac{1}{2v} \int d { \bf r} 
\big (  \omega_p- \frac{ z_p}{z_c}f \omega_c \big )^2 \nonumber \\
&&+\frac{1}{8 \pi l_B z_c^2 }  | \nabla \omega_c|^2 - \frac{n_p}{V} 
\ln { \cal Q}_p [ i \omega_p]- \frac{n_c}{V} \ln { \cal Q}_c [ i \omega_c] \\ \nonumber
\end{eqnarray}
and where the partition function, 
\begin{equation}
{ \cal Z}_0 = \exp \{ \bar{ \phi}_p \ln \bar{ \phi}_p+\bar{ \phi}_c 
\ln \bar{ \phi}_c+\mu_{0p} V \bar{ \phi}_p+ \mu_{0c}V \bar{ \phi}_c \},
 \end{equation}
also written as 
\begin{equation}
{ \cal Z}_0 = \frac{ \zeta_p^{n_p} \zeta_c^{n_c}}{n_p! n_c!} e^{ \beta/2W_{pp}}, 
\end{equation}
is the ideal non-interacting partition function of a mixture of 
$n_p$ charged polymers and $n_c$ counter-ions in solution.
Changing variables we write,
\begin{eqnarray}
&&\beta { \cal H} [ \omega,\omega_c ]= \int d { \bf r} \frac{1}{2 v}  
\omega^2+\frac{1}{8 \pi l_B z_c^2 }  | \nabla \omega_c|^2 \nonumber \\
&& - n_p \ln { \cal Q}_p[ i \omega + i \frac{z_p}{z_c}f \omega_c ]- n_c 
\ln { \cal Q}_c [ i \omega_c]. \\ \nonumber
\end{eqnarray}
The saddle point free-energy is 
\begin{eqnarray}
&&\frac{F}{k_BT}= \beta { \cal H}[ -i w ,-i \psi ]/ \rho_0= \frac{1}{ 8 \pi l_B}
\int d { \bf r} |\nabla \psi( { \bf r})|^2  \nonumber \\
&&-\frac{1}{2 v }  \int d{ \bf r} w^2( { \bf r})- n_p \ln { \cal Q}_p 
\big [ w_p+ \bar{z}_p \psi \big ]-n_c \ln { \cal Q}_c,~~ \\ \nonumber  
\end{eqnarray}
where $v$ is the excluded-volume parameter and where $l_B= e^2/k_BT 
\varepsilon$ is the Bjerrum length.
Note that the electrostatic interaction energy term can be rewritten, 
using equation (\ref{eff}), 
\begin{eqnarray}
\label{free_energy}
&&\frac{F}{k_BT}= - \frac{1}{2 v} \int d { \bf r} w^2( { \bf r})- \frac{1}{2} 
\int d { \bf r} \psi( { \bf r}) \big ( \phi_p( { \bf r})-\phi_c ( { \bf r}) 
\big ) \nonumber \\
&& ~~~~~~~~~~- n_p \ln { \cal Q}_p \big [ w_p + z_p f \psi \big ]-n_c 
\ln { \cal Q}_c \big [  \psi \big ]. \\  \nonumber
\end{eqnarray}
Equation (\ref{free_energy}), after a few steps can be shown to 
reduce to the free-energy expression discussed in equation (\ref{fe}).

\section{ \label{APP:C} On Abel-Volterra equations of the second kind}
\renewcommand{\theequation}{{}\thesection\arabic{equation}}
\setcounter{equation}{0}
In this section we give a detailed derivation of the integral 
Abel-Volterra equations of the first and second kind discussed in section 
\ref{SectionVI}. Consider the integral equation

\begin{equation}
\int_{ U}^{U_0} dt f(t) (t-U)^{-1/2}=g(U).
\label{Abel1}
\end{equation}
We write \cite{Netz},
\begin{eqnarray}
&& ~~~~~~~~~~\int_{ \eta}^{U} dU_0 g(U_0)(U_0- \eta)^{-1/2}= \nonumber \\
&& \int_{ \eta}^{U}
 dU_0 (U_0- \eta)^{-1/2} \int_{ U}^{U_0} dt f(t) (t-U)^{-1/2}. \\ \nonumber
\end{eqnarray}
Using Fubini's theorem we find,
\begin{eqnarray}
\label{Abel2}
&&~~~~~~~~~\int_{ \eta}^{U_0} dU g(U) (U- \eta)^{-1/2}= \nonumber \\ 
&&\int_{ \eta}^{U_0} dt f(t) 
\int_{ \eta}^{t} dU (U - \eta)^{-1/2} (t-U)^{-1/2}. \\ \nonumber
\end{eqnarray}
According to the assumption of strong-stretching theory, one neglects 
the chain end tension $V_0 \approx 0$; the parabolic form in equation (\ref{sst}) follows considering the constraint in equation (\ref{isochrone2}), corresponding to $g(U)=1$ in equation (\ref{Abel1}). This yields
\begin{equation}
f( \eta)= \frac{1}{ \pi} (U_0- \eta)^{-1/2}, 
\end{equation}
that can be integrated to obtain the parabolic form in equation (\ref{sst2}).
Similarly, an expression for the end-monomer distribution can be obtained 
considering 
\begin{equation}
 \int_Z^H dZ_0 ~ g(Z_0) ~ (Z_0^2-Z^2)^{-1/2}= \frac{\pi}{2} \phi_p(Z)
\end{equation}
and inverting this expression \cite{Netz,Zhulina} by fractional 
differentiation, 
\begin{equation}
f(U)= - \frac{1}{ \pi} \frac{ d}{ dU} \int_{U}^{U_0} dt ~ g(t) ~(t-U)^{-1/2},
\end{equation}
one finds equation (\ref{gz}). As discussed in section \ref{SectionVI}, 
let us consider now the case of a uniform tension at the chain ends.
The integral form in equation (\ref{Abel2}) can be written as 
\begin{eqnarray}
&&~~~~~~~~~~~~~\int_{ \eta}^{U_0} dU (U- \eta)^{-1/2}= \nonumber \\
&& \int_{ \eta}^{U_0} dt f(t) 
 \Big \{ \frac{ \pi}{2}+ \arcsin \Big ( \frac{ (t-\eta)-V_0^2}{(t-\eta)+V_0^2} 
\Big ) \Big \}. \\ \nonumber
\end{eqnarray}
Expanding to leading order yields an Abel Integral equation of the 
second kind,
\begin{equation}
\int_{ \eta}^{U_0} dU (U- \eta)^{-1/2}= \int_{ \eta}^{U_0} dt f(t) 
 \Big \{ \pi- \alpha (t-\eta)^{-1/2} \Big \},
\end{equation}
where $ \alpha = \sqrt{6} |V_0|$ measures the uniform end-monomer tension induced
by the electric field, as discussed in section \ref{SectionVI}. We find:

\begin{eqnarray}
\label{Abel3}
f( \eta)&+& \frac{ \alpha}{ \pi} \frac{ d}{ d \eta} \int_{ \eta}^{U_0} dt f(t) 
( t- \eta)^{-1/2} \nonumber \\
=&-& \frac{1}{ \pi} \frac{ d}{ d \eta} \int_{ \eta}^{U_0} dt 
(t - \eta)^{-1/2}. \\ \nonumber
\end{eqnarray}
This can be written as
\begin{equation}
f( \eta)= - \frac{1}{\pi} \frac{d}{d \eta} \int_{ \eta}^{U_0} \big (1+ \alpha f(t) \big)
(t- \eta)^{-1/2}dt,
\end{equation}
and
\begin{equation}
\int_{ \eta}^{U_0} f(t)(t- \eta)^{-1/2} dt = 
 \alpha f( \eta) - 1.
\label{Abel4}
\end{equation}
The presence of the external field converts the Abel-Volterra 
integral equation of the first kind (\ref{Abel1}), as usually discussed 
within strong-stretching theory, to an integral equation of the second 
kind (\ref{Abel3}).
The integral equation (\ref{Abel3}) can be solved and a new form of the 
potential follows, as we will show in the rest of this Appendix.
To leading order in the Taylor expansion, we find
\begin{eqnarray}
&&~~~~~~~~~\frac{df( \eta)}{d \eta} - \frac{ \pi}{\alpha^2} f( \eta) 
=  \frac{ d}{ d \eta} F( \eta) \\ \nonumber
&&F( \eta)= g( \eta)- \frac{1}{ \alpha} \int_{ \eta}^{U_0} dt g( \eta) 
(t- \eta)^{-1/2}.\\ \nonumber 
\end{eqnarray}
The isochronicity constraint in equation (\ref{isochrone}) requires the function 
$g( \eta)$ to be constant, namely $g( \eta)=-1/ \alpha$, so that 
\begin{equation}
F( \eta) = - \frac{1}{ \alpha} + \frac{2}{ \alpha^2} (U_0-\eta)^{1/2},
\end{equation}
and the solution to the Abel-Volterra integral equation of the second kind 
(\ref{Abel3}) has the form
\begin{equation}
f( \eta)= F( \eta)+ \frac{ \pi}{ \alpha^2} \int_{ \eta}^{U_0} dt 
e^{ \frac{\pi}{ \alpha^2}(U_0-t)}F(t)
\end{equation}
and after a few steps we find
\begin{equation}
f( \eta)= \frac{1}{ \alpha} e^{ \frac{\pi}{ \alpha^2}(U_0- \eta)} 
\frac{1}{ \sqrt{\pi}} \Gamma \Big( \frac{1}{2},\frac{\pi}{ \alpha^2}
(U_0- \eta) \Big),
\label{SST2}
\end{equation}
where $\Gamma( \alpha,x)$ is the incomplete gamma function of order $1/2$ 
and where the potential form can be obtained integrating equation (\ref{SST2}).
We note that the function $f(U)= \sqrt{\frac{3}{2}} dZ/dU$ can be written as
\begin{equation}
f( \eta)= \frac{1}{ \alpha \sqrt{ \pi}}~{ \cal U} 
\Big(\frac{1}{2},\frac{1}{2},\frac{\pi}{\alpha^2}(U_0- \eta) \Big),
\end{equation}
where ${ \cal U} \Big(\frac{1}{2},\frac{1}{2},\frac{\pi}{\alpha^2}(U_0- \eta) 
\Big)$ is the confluent hyper-geometric Tricomi function and 
where one recovers the standard strong-stretching theory prediction to first order 
expansion. 
It is easily checked that 
the first order in the above mentioned expansion corresponds to the strong-stretching 
theory result:
\begin{equation}
f( \eta)= \frac{1}{ \pi} ( U_0 - \eta)^{-1/2},
\end{equation}
that corresponds to the parabolic form (\ref{sst2}). Equation 
(\ref{SST2}) can be integrated, and one finds
\begin{equation}
\sqrt{ \frac{3}{2}} \pi Z = \frac{ \alpha}{ \sqrt{ \pi}} 
\int_0^{ \frac{ \pi}{ \alpha^2}(U_0-U)}  \Gamma\big ( \frac{1}{2},t \big ) 
e^t  ~dt, 
\end{equation} 
and finally 
\begin{eqnarray}
&&\sqrt{ \frac{3}{2}} \pi Z = 2 (U_0-U)^{1/2}+ \alpha - 
\frac{ \alpha}{ \sqrt{ \pi}}
e^{ \frac{ \pi}{ \alpha^2} ( U_0-U)} \nonumber \\
&& ~~~~~~~~~~~~~\times \Gamma 
\big ( \frac{1}{2}, \frac{ \pi}{ \alpha^2}(U_0-U) \big ).
\end{eqnarray}
The above expression has been used to obtain the potential, for both negative 
and positive values of the chain-end force, shown in Fig.~\ref{fig11} and 
discussed in section \ref{SectionVI}.
For the chain end distribution we find 
\begin{eqnarray}
\frac{ds( \eta)}{d \eta} &-& \frac{ \pi}{\alpha^2} s( \eta) 
=  \frac{ d}{ d \eta} G( \eta) \\ \nonumber
G( \eta) &=& -\frac{1}{ \alpha} \eta + \frac{4}{ 3 \alpha^2} \eta^{3/2}, \\ \nonumber 
\end{eqnarray}
where $s( \eta)=g( \eta) \frac{dZ}{ d \eta}$. The expression for the chain-end distribution in equation (\ref{SST2g}) can be obtained solving the above second kind Abel-Volterra integral 
equation as well as using the result of equation (\ref{SST2}) above. 

.
\end{document}